# Pressure tuning of hydrogen bond ordering in the metal-organic framework [(CH$_3$)$_2$NH$_2$]Mn(HCOO)$_3$

Na Su [1,2], Yinina Ma [2,3], Shuang Liu[1], Wei Wu [2,3], Jianlin Luo[2,3,*], and Young Sun[1,*]

[1] Department of Applied Physics and Center of Quantum Materials and Devices, Chongqing University, Chongqing 401331, China

[2] Beijing National Laboratory for Condensed Matter Physics and Beijing Advanced Innovation Center for Materials Genome Engineering, Institute of Physics, Chinese Academy of Sciences, Beijing 100190, China

[3] School of Physical Sciences, University of Chinese Academy of Sciences, Beijing 100190, China

*Corresponding authors: jlluo@iphy.ac.cn; youngsun@iphy.ac.cn

Abstract

The influence of pressure on the hydrogen bond ordering in the perovskite metal-organic framework [(CH$_3$)$_2$NH$_2$]Mn(HCOO)$_3$ has been investigated by dielectric, pyroelectric adn magnetic measurements in a piston-cylinder cell. Under ambient pressure the ordering of hydrogen bonds takes place at $T_C$ = 188 K and induces a first-order ferroelectric phase transition. With increasing pressure to p = 3.92 kbar, the order-disorder transition shifts to a lower temperature and retains the first-order ferroelectric nature. However, under higher pressures, the ordering process of hydrogen bonds is split into two transitions: a broad antiferroelectric transition at high temperature and a first-order ferroelectric transition at low temperature. With increasing pressure, the antiferroelectric phase is enhanced whereas the ferroelectric phase is greatly suppressed, which implies that compression of the perovskite framework favors antiparallel arrangement of the hydrogen bonds. The canted anti-ferromagnetic transition was almost unchanged when pressure up to 10.85 kbar. Our study demonstrated that the perovskite metal-organic frameworks are more sensitive to external pressure than conventional perovskite oxides so that their electric properties can be easily tuned by pressure.

## I. INTRODUCTION

Metal-organic frameworks (MOFs) that combine inorganic and organic elements have been extensively studied in the past decade due to their potential applications in gas storage, catalysis as well as rich physics in optics, magnetism, and ferroelectricity [1-5]. Particularly, a lot of attention has been focused on the dense MOFs where electric and magnetic orders can coexist [6-21]. Among them, a metal formate family with the general formula of $[(CH_3)_2NH_2]M(HCOO)_3$ ($M$ = Mn, Co, Ni, Fe), termed as DMA-$M$, has attracted a wide attention as a new group of multiferroic materials [7]. Their structures are isomorphic, belonging to the $ABX_3$ perovskite architecture, where metal cations ($B$) connected by formate groups ($X$) constitute the framework and the DMA cations ($A$) are located in cavities, as shown in Fig. S1. The DMA cations form hydrogen bonds with the formate to be stabilized in the cavity. The ordering of hydrogen bonds induces either ferroelectricity or antiferroelectricity with the transition temperature between 190 and 160 K [7-9,11,13-15]. Meanwhile, DMA-$M$ also shows ferromagnetic or canted-antiferromagnetic ordering below 37 K to 9 K [8,10,14,15,22]. Some of them exhibit obvious magnetoelectric (ME) coupling effect in the paramagnetic state where the electric polarization ($P$) can be modified by applying high magnetic fields[8,9,15]. Moreover, DMA-Fe displays an interesting resonant quantum ME effect in the multiferroic state [11,12].

It is well known that MOFs have relatively soft and flexible structures so that external pressures could deform them easily, which may induce significant modifications in both electric and magnetic properties [23-33]. Previous studies on DMA-$M$s demonstrated that external pressures can induce several transitions in them [32-36]. The room temperature Raman spectra with applied pressures revealed three structural transitions for both DMA-Mn (at 21, 41, and 67 kbar) and DMA-Mg (at 22, 40, and 56 kbar), and two transitions for DMA-Cd (one between 12 and 20 kbar, and another near 36 kbar) [35,36]. However, for DMA-Mn, this Raman result is not inline with the high-pressure X-ray diffraction data reported by Collings et al [33]. They found that DMA-Mn undergoes only one transition near 55 kbar at room temperature during which its structure changes from *R-3c* symmetry into *P$\bar{1}$* symmetry, suggesting antiferroelectric (AFE) order. In addition, calculations on high-pressure magnetic behavior of DMA-Fe predicted a high-spin to low-spin transition near 45 kbar [32]. Although some relevant works have been done, few studies investigated the pressure influence on the order-disorder transition of hydrogen bonds by direct electric measurements.

In this work, we have investigated the pressure effect on the hydrogen bond ordering in DMA-Mn which has the highest transition temperature and the largest electric polarization among the DMA-*M*s family. The temperature dependences of dielectric permittivity and electric polarization under pressures up to 15.19 kbar were measured and a pressure-temperature phase diagram was obtained. It is found that external pressure would suppress FE ordering and favor AFE ordering of hydrogen bonds in DMA-Mn.

## II. EXPERIMENT

The single-crystal samples of $[(CH_3)_2NH_2]Mn(HCOO)_3$ (DMA-Mn) were synthesized by hydrothermal method. 5 mmol $MnCl_2·4H_2O$ was dissolved in a 60 mL solution containing 50% volume of dimethylformamide (DMF) with deionized water. Then,the solution was heated in 100 mL Teflon–lined autoclaves for 3 days at 140°C. After the autoclaves were air cooled, the supernatants were removed into a glass beaker for crystallization by slow evaporation method at room temperature. Finally, the colorless crystals were harvested after 3 days and were washed by ethanol for 3 times. The single-crystal XRD patterns at room temperture suggest that the crystals are naturally grown layer by layer along the [012] direction as shown in Fig. S1.

The electric properties under several pressures were measured with a clamp-type piston cylinder cell [37]. Daphne 7373 was used as the pressure transmitting medium[38], which can maintain a relatively good hydrostatic pressure up to 20 kbar at room temperature. The pressure ($p$) was determined at room temperature from the relative change of superconducting transition temperature of Pb via the formula $p$ (kbar) = $\Delta T/0.0365$ [37].

Temperature dependence of magnetization M(T) under high pressure was measured with a miniature piston-cylinder cell (Quantum Design Japan) in a commercial magnetic property measurement system (MPMS-III) from Quantum Design. The DMA-Mn single crystals together with a piece of Pb was loaded into Teflon capsule filled with Daphne 7373 as the pressure transmitting medium. The magnetic field was applied along the [012] direction. The pressure at low temperatures was determined from the superconducting transition of Pb [37].

The dielectric permittivity was measured with an Andeen Hagerling 2700A capacitance bridge at the frequency of 1kHz, 10 kHz, 20kHz and 100kHz in a PPMS (Quantum Design). In order to obtain the temperature ($T$) dependen electric polarization ($P$), the sample was poled with a 4.2 kV/cm electric field during cooling from a high temperature in the paraelectric phase. After removing the poling electric field, the electrodes were shorted for 40 minutes to release the space charges. Then, the sample was warmed at a rate of 0.5 K/min while the pyroelectric current ($I$) was recorded by an electrometer (Keithley 6517B). Finally, $P$ was obtained by integrating the pyroelectric current with time.

## III. RESULTS and DISCUSSION

Fig. 1 shows the temperature dependence of dielectric permittivity ($\varepsilon_r$) and loss (tanδ) at a frequency of 10 kHz along [012] direction under ambient pressure. On warming up, $\varepsilon_r$ exhibits a step-like anomaly and tanδ shows a sharp peak at $T_C$ = 192 K. The transition temperature during cooling and warming processes displays a hysteresis of 12 K, suggesting a first-order phase transition. As previous studies clarified, during this phase transition, the crystal structure of DMA-Mn transforms from a centrosymmetric space group *R-3c* to a ferroelectric (FE) space group *Cc*, which is associate with the disorder–order transition of hydrogen bonds[7,8,39].

In order to detect how external pressure affects the ordering process of hydrogen bonds in DMA-Mn, we measured the temperature dependence of dielectric permittivity and electric polarization under various pressures.

As shown in Fig. 2(a-C), 0 kbar means the sample was put in the piston cell without pressure. The dielectric anomaly and the peak in tanδ shift slightly to a lower temperature ($T_C$ = 191.5 K) at 0 kbar, which means the structure of DMA-Mn is much sensitive to weak pressure. With further increasing pressure ($p$ = 3.92 kbar), the transition temperature shifts to a lower temperature ($T_C$ = 175 K). The sharp phase transition as well as the broad hysteresis between the warming and cooling processes suggest that the order-disorder process of hydrogen bonds remains the first-order nature in the low pressure range.

Fig. 3 and Fig. 4 present the dielectric behaviors at a frequency of 10 kHz under higher pressures up to 15.19 kbar. With increasing pressure, the sharp phase transition is gradually broadened and split into two transitions. The transition temperatures ($T_1$ and $T_2$) can be identified based on the peak/hump in tanδ or εr. $T_1$ decreases rapidly toward low temperatures as the applied pressure grows whereas $T_2$ is only slightly dependent on pressure. Meanwhile, the thermal hysteresis is greatly reduced under high pressures. The dielectric behaviors of multifrequency on warming process under pressure were shown in Fig. S2 and Fig. S3. $T_2$ has a strong frequency-dependent behaviour, which occurs due to the relaxation of the electric dipole moment and slow dynamics of electric dipole to ac electric field stimuli. These results indicate that the collective ordering of hydrogen bonds is driven into two continuous partial ordering by pressure.

The nature of the two successive transitions is revealed by the measurements of temperature dependence of pyroelectric current and electric polarization. As shown in Fig. 5, under various pressures up to 12.95 kbar, the polarization appears around $T_1$, but no signal exists around $T_2$. This implies that the phase transition at $T_1$ is ferroelectric and that at $T_2$ could be relaxation-type antiferroelectric. The intensity of polarization around $T_1$ decays rapidly with increasing pressure, and becomes undetectable at 15.19 kbar. Accordingly, the electric polarization below $T_1$ are greatly suppressed by pressure. Therefore, it is concluded that external pressure destabilizes the ferroelectric phase but favors the antiferroelectric phase.

The magnetization curves and it's derivative curves of DMA-Mn with the applied magnetic field of 0.1 T under pressure are shown in Fig. 6. The sharp peaks of derivative of the magnetization curve signify the canted anti-ferromagnetism. To our surprise, this transition temperature monotonically increases slightly with pressure and can be considered almost unchanged. When the pressure increases to 10.85 kbar, $T_{AFM}$ only increases from 8.4 K to 9 K, which is almost negligible compared to the pressure regulated ferroelectric transition temperature. The long-range super-exchange model Mn-O-C-O-Mn determines the magnetic order of DMA-Mn, pressure has little impact on the framework according to the Goodenough-Kanamori rules[40].

A temperature-pressure phase diagram of DMA-Mn can be constructed based on the dielectric, pyroelectric and magnetic data under pressures. At high temperatures, DMA-Mn exhibits the PE behavior due to the disorder of hydrogen bonds. Under zero or low pressures, the disordered hydrogen bonds become collectively parallel alignment upon cooling down. As a result, a sharp first-order phase transition from the PE to FE phase occurs. However, external pressure destabilizes the parallel alignment of hydrogen bonds and favors antiparallel ordering. With decreasing temperature, partial hydrogen bonds become antiparallel alignment, which yields the AFE phase at $T_2$. With further cooling, the remaining disordered hydrogen bonds develop into the parallel ordering, which makes the AFE and FE phases coexist below $T_1$. Under a high pressure of 15.19 kbar, the FE phase completely disappears and only the AFE phase remains.

Previous studies of the Raman spectra and single-crystal X-ray diffraction under pressures suggest that the metal-formate framework of DMA-Ms is much more susceptible to external pressures than the hydrogen bonds [33,35,36]. Moreover, the neutron scattering data of deuterated DMA-Mn imply that the framework distortion plays an important role in the arrangement of hydrogen bonds [41]. Here, we demonstrated the framework was more robust to the pressure than hydrogen bonds, but slight framework distortion will also affect the hydrogen bonds. The slightly anisotropic distortion of the $MnO_6$ octahedra under pressures reduces the size of the framework cavities and changing their shapes. The smaller cavities make the dipolar interaction between the DMA cations strengthened gradually so that more and more hydrogen bonds prefer to orient in antiparallel to reduce the total energy. Thus, with increasing pressure, the FE phase is gradually suppressed and the AFE phase is strengthened.

## IV. CONCLUSION

In summary, our study reveals that the ordering process of hydrogen bonds in the perovskite MOF, DMA-Mn is very sensitive to external pressure. Under weak pressure, a sharp first-order phase transition from the PE to FE state is induced by the collective alignment of hydrogen bonds. When applying a relatively high pressure, the ordering process of hydrogen bonds is broadened and split into two transitions, which results in a coexistence of the FE and AFE phases. The FE phase is completely suppressed under 15.19 kbar but the AFE state is stabilized. It is concluded that external pressure favors the antiparallel ordering of hydrogen bonds so that the FE state decays rapidly with increasing pressure. These results demonstrate that pressure is an effective tool to tune the physical properties of MOFs because of their relatively flexible lattice.

This work was supported by the National Natural Science Foundation of China (Grant No. 12227806), the National Key Research and Development Program of China (Grant No. 2021YFA1400303), and the Fundamental Research Funds for the Central Universities (Grant No. 2023CDJXY-0049).

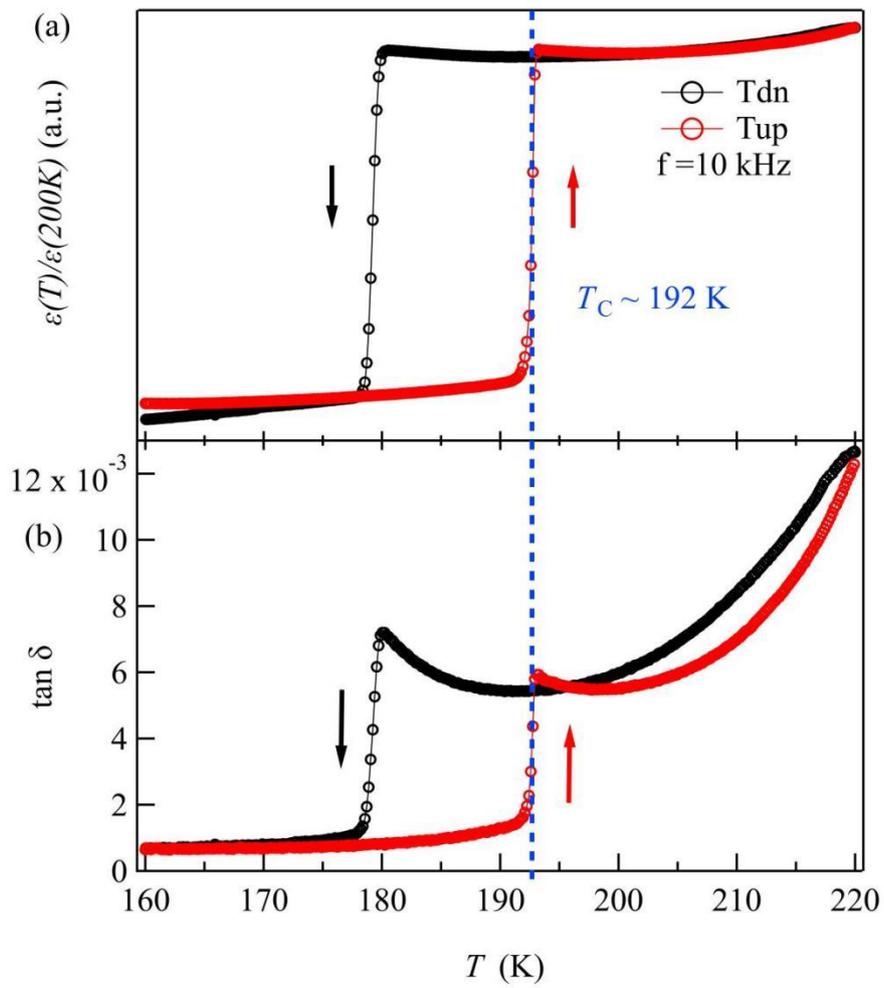

**Figure 1** The temperature dependence of normalized dielectric permittivity ($\varepsilon/\varepsilon_0$) (a) and loss (tan$\delta$) (b) along [012] direction under ambient pressure during cooling and warming process.

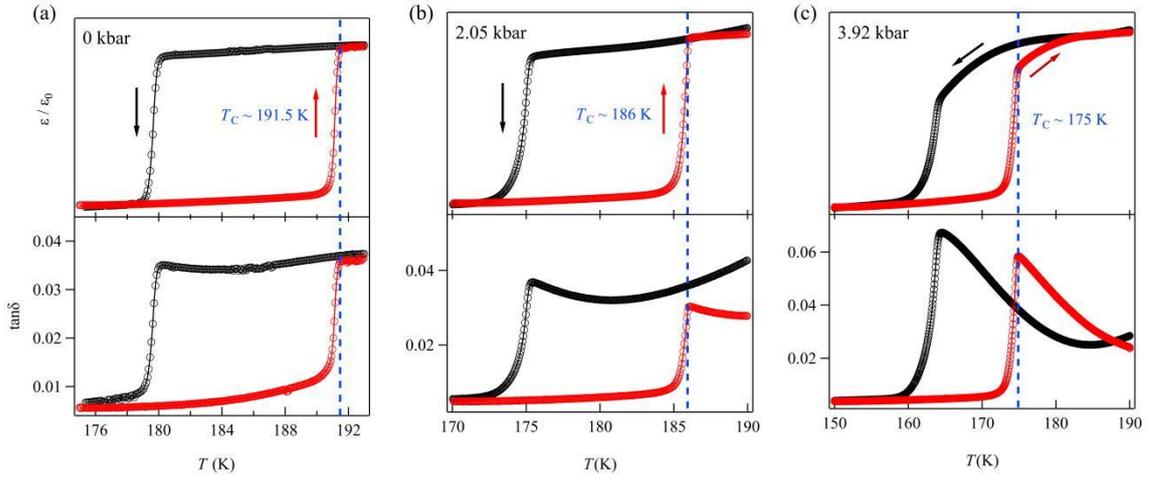

**Figure 2** The normalized dielectric permittivity ($\varepsilon/\varepsilon_0$) and dielectric loss (tan$\delta$) as a function of temperature along [012] direction under (a) 0 kbar, (b) 2.05 kbar and (c) 3.92 kbar.

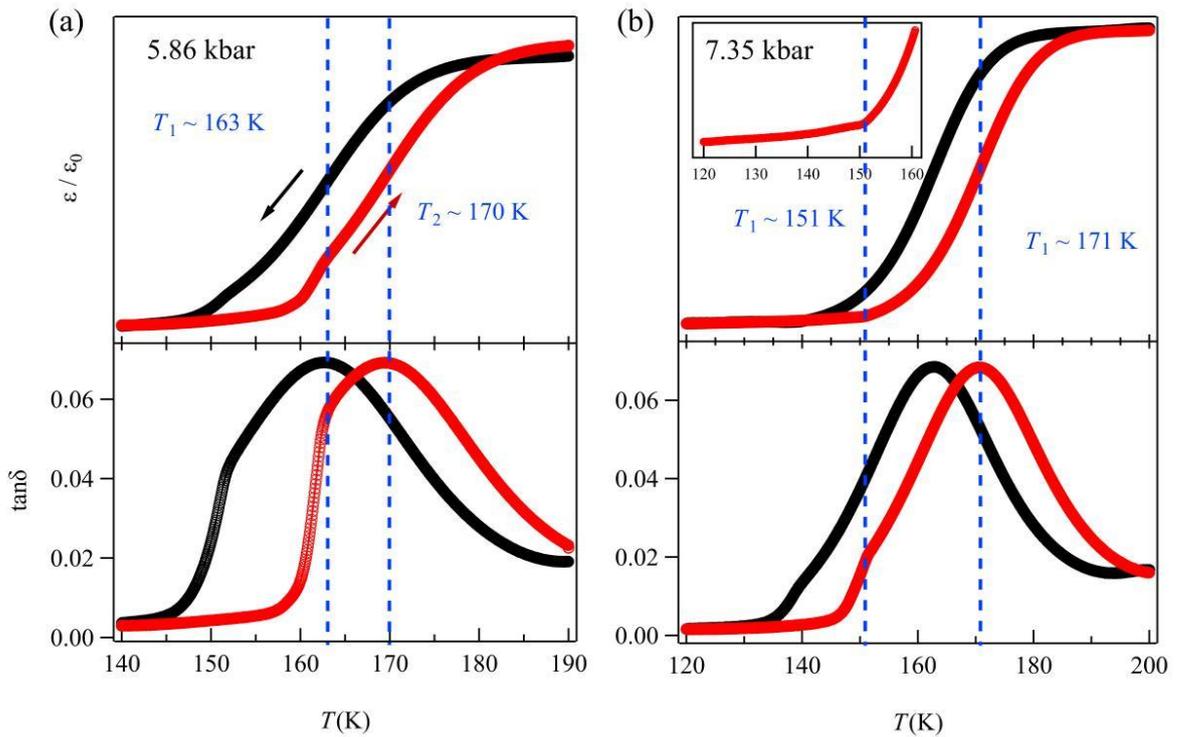

**Figure 3** The normalized dielectric permittivity ($\varepsilon/\varepsilon_0$) and dielectric loss (tan$\delta$) as a function of temperature along [012] direction under (a) 5.86 kbar and (b) 7.35 kbar. The inset in (b) shows the enlarged view around $T_1$.

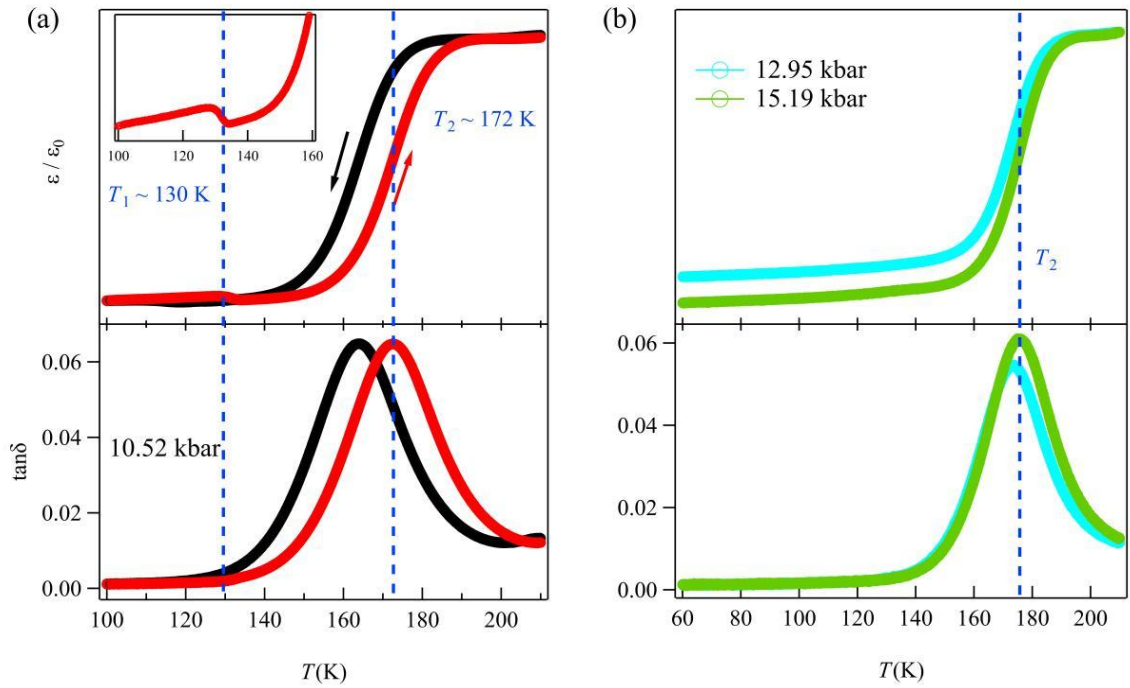

**Figure 4** (a) The normalized dielectric permittivity ($\varepsilon/\varepsilon_0$) and dielectric loss (tan$\delta$) as a function of temperature along [012] direction under 10.52 kbar, the inset shows the dielectric anomaly in the low temperature range. (b) he normalized dielectric permittivity ($\varepsilon/\varepsilon_0$) and dielectric loss (tan$\delta$) as a function of temperature along [012] direction under 12.95 kbar and 15.19 kbar.

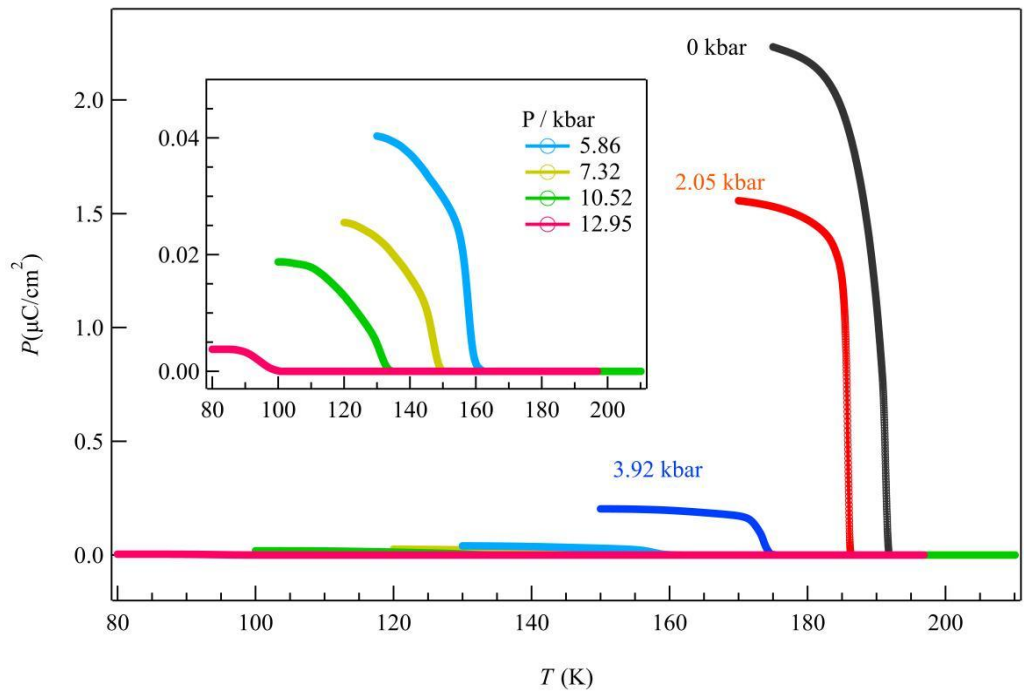

**Figure 5** The temperature dependence of electric polarization (*P*) along the [012] direction under different pressures. The inset shows the enlarged view under high pressures.

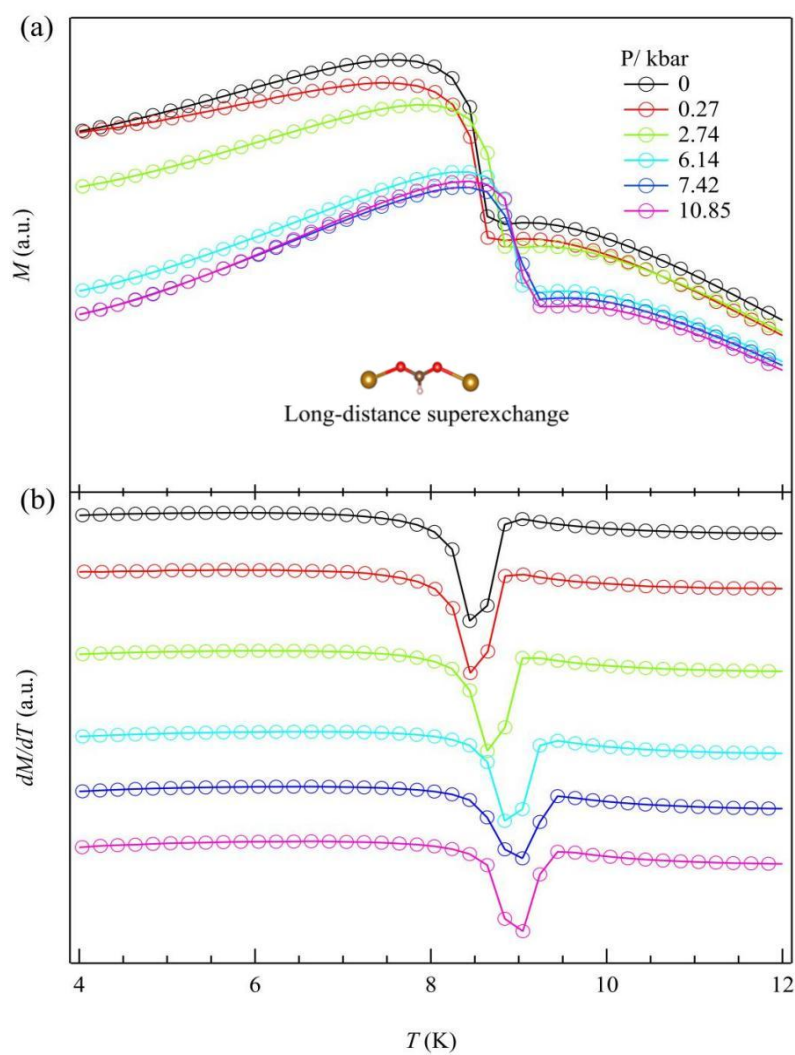

**Figure 6** (a)The temperature dependence of magnetization along the [012] direction under different pressures. (b) The derivative of the temperature-dependent magnetization under pressures along the [012] direction.

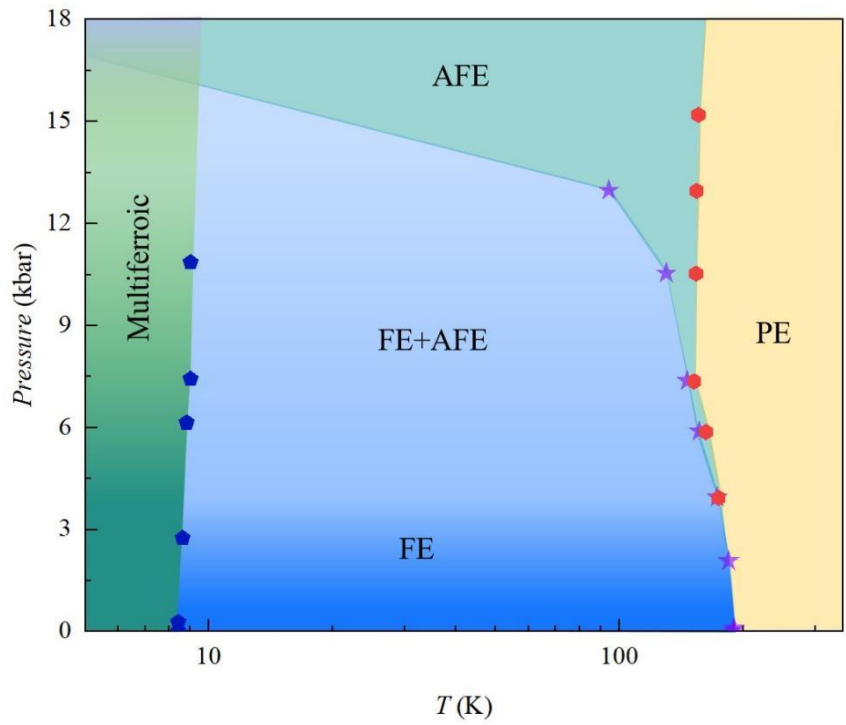

**Figure 7** The pressure-temperature phase diagram of DMA-Mn.